\documentclass[default,pdflatex]{sn-jnl}
\usepackage[utf8]{inputenc}	
\usepackage[english]{babel}	
\usepackage{2022-Computing-Cruz}

\jyear{2022}

\begin{document}

\title[\SHORT]{\TITLE}

\author*[1,2]{\fnm{Margarita} \sur{Cruz}}\email{cruz@us.es}
\author[1,2]{\fnm{Beatriz} \sur{Bernárdez}}\email{beat@us.es}
\author[1,2]{\fnm{Amador} \sur{Durán}}\email{amador@us.es}
\author[3]{\fnm{Cathy} \sur{Guevara-Vega}}\email{cguevara@utn.edu.ec}
\author[1,2]{\fnm{Antonio} \sur{Ruiz-Cortés}}\email{aruiz@us.es}

\affil[1]{%
  \orgdiv{\ITRESUS}, %
  \orgname{\US}, %
  \orgaddress{%
    \city{\Sevilla}, %
    \country{\Spain}%
  }%
}

\affil[2]{%
  \orgdiv{\SCORE}, %
  \orgname{\US}, %
  \orgaddress{%
    \city{\Sevilla}, %
    \country{\Spain}%
  }%
}

\affil[3]{%
  \orgdiv{eCIER Research Group}, %
  \orgname{Universidad Técnica del Norte}, %
  \orgaddress{%
    \city{Ibarra}, %
    \country{Ecuador}%
  }%
}





\newcommand{\gls}[1]{\textcolor{red}{#1}}

%
%
%
%


\newcommand{\Science}{Science\xspace}
\newcommand{\Agrobiology}{Agrobiology\xspace}
\newcommand{\Agrobio}{Agrobio\xspace}
\newcommand{\Soil}{Soil\xspace}
\newcommand{\Harvest}{Harvest\xspace}
\newcommand{\Bio}{Bio\xspace}
\newcommand{\Olive}{Olive\xspace}
\newcommand{\Diet}{Diet\xspace}

\newcommand{\Automatic}{Automatic\xspace}
\newcommand{\Computational}{Computational\xspace}
\newcommand{\computational}{computational\xspace}
\newcommand{\Comp}{Comp\xspace}
\newcommand{\Testing}{Testing\xspace}
\newcommand{\SPL}{SPL\xspace}

\newcommand{\Expressiveness}{Expressiveness\xspace}
\newcommand{\expressiveness}{expressiveness\xspace}
\newcommand{\Precision}{Precision\xspace}
\newcommand{\precision}{precision\xspace}
\newcommand{\Traceability}{Traceability\xspace}
\newcommand{\traceability}{traceability\xspace}
\newcommand{\Comprehension}{Comprehension\xspace}
\newcommand{\comprehension}{comprehension\xspace}







\abstract{%
\textbf{Context:} %
The need of replicating empirical studies in \CS is widely recognized among the research community as a means to consolidate the acquired knowledge from previous studies and generalizing results. %
%
It is essential to report the changes in the 
settings of each replication to promote not only the comprehensibility of the evolution of the experimental validity of an original study across a family of studies, but also replicability itself. %
Unfortunately, the lack of proposals for the systematic reporting of changes in replication undermines these desirable objectives. %

%
\textbf{Objective}. %
The main goal of the work presented in this article is to provide researchers in \CS, and in other areas of research when appropriate, with a systematic, tool-supported approach for the specification and reporting of changes in the replications of their empirical studies. %
\textbf{Method:} %
Applying \emph{Design Science Research}, once the problem was identified, we have developed and validated a composite artifact consisting of (i) a metamodel that formalizes all the relevant concepts related to replications and their changes; (ii) templates and linguistic patterns that facilitates the reporting of those relevant concepts; and (iii) a proof-of-concept model-based software tool that supports the proposed approach. %
For its validation, we have carried out a multiple case study including 9 families of empirical studies not only from \CS, but also from an area as different as \Agrobiology, in order to check the external validity of our approach when applied to research areas for which it was not initially designed. 
The 9 families encompass 23 replication studies and a total of 92 replication changes, for which we have analyzed the suitability %
of our proposal. %
\textbf{Results:} %
The multiple case study revealed some initial limitations of our approach, such as shortcomings related to threats to experimental validity or context variables. After several improvement iterations on the artifact, all the 92 replication changes could be properly specified, %
including also their qualitatively estimated effects on the different types of experimental validity across the entire family of experiments and its corresponding visualization. %
%
\textbf{Conclusions:} %
Our proposal for the specification of replication changes seems to fit the needs not only of replications in \CS, but also in other research areas. 
Nevertheless, further research is needed to improve it 
and to disseminate its use among the research community.
}


\maketitle



\section{Introduction} \label{sec:intro}

%
As in most research areas, empirical studies, especially controlled experiments, can be used in \CS to rigorously evaluate technologies, methods, and tools and help guide further research by revealing existing problems and difficulties \cite{ciolkowski2002family}. However, for their results to be generalizable, reported experiments must be replicated in different contexts, at different times, and under different conditions \cite{campbell2015experimental} by means of so-called \emph{families of experiments}.\footnote{The concept of \emph{family of experiments} can be understood as an specialization of the more general concept of \emph{family of empirical studies}. Since experiments are the most common type of empirical study in \CS and its related disciplines, in this article we will use mainly the term \emph{family of experiments}, but most of the proposal is also applicable to other types of replicable empirical studies such as, for example, case studies or surveys.} %
As \citeauthor{basili1999building} introduced in \citeyear{basili1999building} \cite{basili1999building}, a family of experiments consists of a \emph{baseline} or \emph{original study}, followed up by a set of replications that answer the same research questions as the original study. %
Later, \citeauthor{santos2018analyzing} in \citeyear{santos2018analyzing} \cite{santos2018analyzing}, proposed the following premises to consider a series of experiments as a family, namely: (i) access to the raw data is guaranteed; (ii) researchers know the exact setup of each experiment; and (iii) at least three experiments evaluate the effects of at least two different technologies (or methods or tools) on the same response variable. %
Nowadays, it is widely accepted in the research community that the knowledge obtained from a family of experiments is more robust and reliable than that obtained from a single isolated experiment, the results of which can only be considered as preliminary \cite{shull2008role,spence2016prediction}. %

Let us consider for a moment a novice researcher in \CS who decides to replicate an original study or a previous replication of an original study. %
According to \citeauthor{carver2010towards} \cite{carver2010towards}, she needs to carefully review the entire family of previous studies, in order to acquire the necessary knowledge to properly adapt the experimental settings, improve its design---if possible---to increase its experimental validity, or avoid making the same mistakes than in previous studies. %
To alleviate this situation, several initiatives have recently emerged, the most relevant being \emph{Open Science} \cite{mendez2020open}, which promulgates that making available the datasets, the analyses, and a \emph{preprint} version of an experiment and its related software, provides valuable knowledge allowing not only that any interested party may audit it, but also that others build directly upon the previous work through reuse and repurposing \cite{ACM-2020}. %

A complementary approach to increase the visibility and reproducibility of empirical studies in \CS 
is promoting the availability of artifacts such as \emph{laboratory} or \emph{replication packages} \cite{basili1999building,solari2006classifying}. %
According to \cite{shull2008role}, replication packages should include not only datasets, analyses, and experimental material, but also guidelines to conduct a replication and a summary of the evolution of the experiment across the family \cite{solari2017content}, promoting traceability among replications. %

Despite all these efforts, \citeauthor{shepperd2018role} \cite{shepperd2018role} stated that although there is a consensus on replications being essential to consolidate the findings of empirical studies, there is a need for better reporting of both original and replicated studies. %
To report a replication of a controlled experiment, \CS researchers usually use as a reference some seminal works such as those by \citeauthor{wohlin:experimentation} \cite{wohlin:experimentation}, \citeauthor{jedlitschka2008reporting} \cite{jedlitschka2008reporting}, or \citeauthor{juristo2013basics} \cite{juristo2013basics}, complementing them with \citeauthor{carver2010towards}'s replications guidelines \cite{carver2010towards}. %
The recommendations in the aforementioned works are meant for controlled experiments in general, but not for the specification of the changes that often arise during replications to address threats to experimental validity and, therefore, improve the original study and the validity of its results \cite{Brooks2008}. %

To the best of our knowledge, there is a lack of specific proposals to specify the changes introduced between replications of the same family of experiments. In this situation, researchers choose either to report the experimental setting of the replication without highlighting the changes from the original study or previous replications \cite{herbold2017global,mondal2018cloned,santos2019comparing,fernandez2016does}, or they report the changes in an ad-hoc manner, describing them in narrative text \cite{reimanis2014replication,assar2016using,albayrak2014investigation,riaz2017identifying}. %
This lack of detail in the specification of changes leads to some problems in carrying out new replications, since the replicator has difficulties not only in deciding which aspects of the experimental setup are best suited to the new environment, but also in avoiding mistakes made in previous experiments that threatened their validity \cite{carver2010towards,ACM-2020}. %
On the other hand, a proper knowledge of the changes allows a better meta--analysis of the family, since they impact on issues such as the definition of moderator variables, the definition of the aggregated family design, or the choice of the most appropriate analysis, among others \cite{bernardez2020effects}. %


The proposal described in this article aims to specify replication changes in a structured, systematic way, identifying aspects such as the rationale for each change or its effect on experimental validity, which can be quantified and used to visualize the evolution of the entire family of experiments, thus supporting decision making for new replications \cite{carver2010towards}. %
By specifying changes explicitly, our proposal helps to decrease not only the so-called \emph{tacit knowledge} \cite{shull2002replicating}, but also \emph{experimenter bias} \cite{dos2022investigating}, that occurs when replicators have to interact with the original experimenters to request missing information about the family, specifically their changes. %

In this work, we focus on the specifications of replications and their corresponding changes of empirical studies in general and controlled experiments in particular. For that purpose, we have adopted \emph{Design Science Research} (\DSR) as our research methodology, creating and evaluating an artifact that is designed to solve an identified problem \cite{von2004design}. %
In our case, the developed artifact consists of (i) a metamodel developed by an iterative and incremental process that represents the relevant concepts about replications and their changes; (ii) templates for reporting the information included in the metamodel, thus promoting reusability, avoiding redundancies, and tracing the effects of changes on experimental validity across the family of experiments; and, (iii) \caesar, a proof-of-concept model-driven software tool developed to provide an initial evaluation and support for our proposal, including the aforementioned templates and the automatic visualization of the effect of the changes on experimental validity across the family of experiments.

The rest of the paper is organized as follows. %
Section \ref{sec:background} 
summarizes a brief state-of-the-art on families of experiments, replications, and changes in \CS. %
Section \ref{sec:metamodelo} 
presents the three components of the developed artifact, i.e. the metamodel, the templates, and the tool support. %
Section \ref{sec:case_study} 
reports a multiple case study carried out to evaluate the artifact, encompassing 9 families of experiments from \CS and 
\Agrobiology. %
%
%
Related work is commented in Section \ref{sec:trabajos} 
and, finally, concluding remarks and future work are presented in Section \ref{sec:conclusions}. 
    


\section{Background} \label{sec:background}

According to \cite{1330459}, a replication can be defined as the repetition of an experiment, either as closely following the original experiment as possible, or with deliberate changes to the original experiment’s settings, in order to achieve, or ensure, greater validity in the carried out research. 
Despite the importance of replications, and although their practice has increased in recent years \cite{da2014replication,cruz2019replication}, the number of replications in \CS in general, and in \SE in particular, remains low \cite{solari2017content}. %
Among the causes influencing this situation are (i) the tacit knowledge not explicit in replication reports \cite{shull2002replicating}; (ii) the lack or incompleteness of laboratory packages \cite{solari2017content}; (iii) the lack of agreement on common terminology and criteria for reporting replications \cite{carver2010towards}; and, last but not least, (iv) the effort and resources needed to carry out an experiment \cite{da2014replication}.

Several taxonomies have been proposed to classify the different types of replications. There is a wide agreement to use the term \emph{internal replication} for those replications carried out by the same experimenters at the same site than the original study, whereas \emph{external replication} is used when the experimental team and site are different from the original ones. %
With respect to the  process to carry out a family of experiments, once the original experiment concludes, it is advisable to carry out internal replications to confirm preliminary results and adjust experimental settings. Then, external replications can be carried out for generalizing the provisional results from internal replications \cite{brooks1996replication}.

Nevertheless, there is a lack of agreement in the used terminology in other replication taxonomies. For example, according to the degree to which the original experiment protocol is followed, \citeauthor{shull2008role} \cite{shull2008role} classify  replications as \emph{exact} and \emph{conceptual}, but \citeauthor{juristo2011role} \cite{juristo2011role} use \emph{closed} and \emph{differentiated} instead. %
\citeauthor{basili1999building} refer to \emph{strict} replications when the original study is duplicated as accurately as possible \cite{basili1999building}, whereas \citeauthor{gomez2014understanding} \cite{gomez2014understanding}, classify them as \emph{literal}, \emph{operational} and \emph{conceptual} depending on the changes carried out and their purpose. %
Other taxonomies such as those proposed in \cite{1330459} and \cite{Baldassarre} use some of the terms commented above.
%
In \cite{de2015investigations}, \citeauthor{de2015investigations} compare different taxonomies, concluding that any attempt to establish a replication typology must be done with care since, as stated in \cite{gomez2010replications}, there are authors who use the same term for different types of replications and conversely, use different terms to refer to the same type of replication. 

In our opinion, the main underlying concept behind the different proposed taxonomies are the changes between the original study and its replications. %
%
%
In particular, in human-oriented experiments such as those common in \SE, each change in the experimental settings, even if the protocol of the original experiment is followed, can eventually produce different results \cite{fernandez2019open}. As a consequence, specifying the changes and their motivation allows the comparison of results and increase knowledge by analyzing the conditions under which the results were obtained, thus encouraging further replication.

Last but not less important, and although several authors \cite{carver2010towards,shepperd2018role} have reported their relevance, only a few replications report their related changes when published. %
When included, changes are reported in narrative text, e.g. \cite{albayrak2014investigation,itkonen2014test,scanniello2015link}, or in a simple, non-standard tabular form, with one row per change including a few properties only, such as elements affected by the change (population, experimental design, etc.) or the situation 
after the change, e.g. \cite{assar2016using,reimanis2014replication,riaz2017identifying}.
\section{Model-based proposal for specifying replication changes} \label{sec:metamodelo} 


In this section, the model-based proposal for the specification of replication changes in 
empirical studies, encompassing a metamodel, some templates, and a proof--of--concept software tool, is presented.

\subsection{Metamodel for Replication Changes} \label{sec:metamodel}

\begin{figure*}
    \centering
    \includegraphics[scale=0.625,trim={1cm 12.50cm 2.25cm 3.00cm},clip]
    {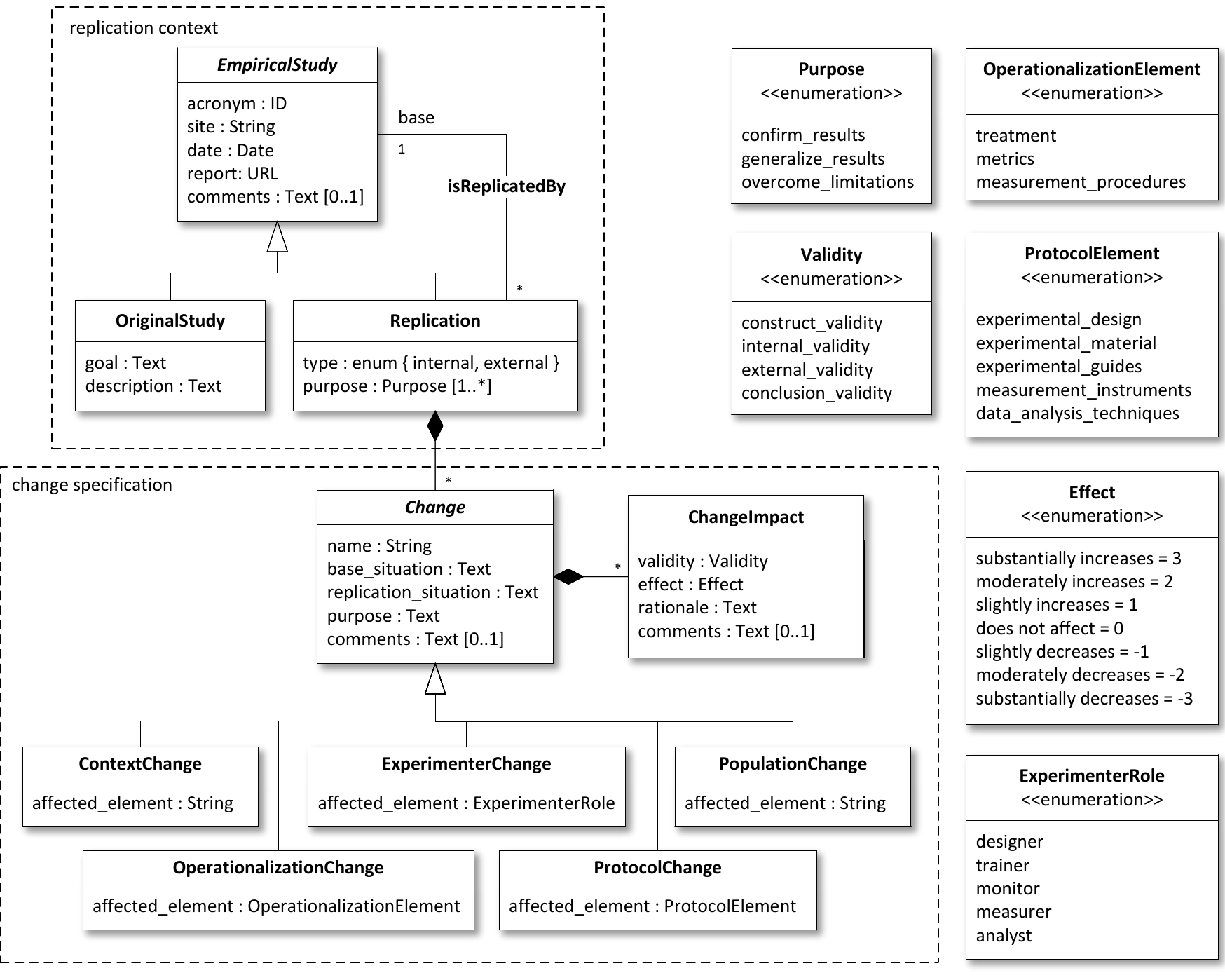}
    \caption{Metamodel for replication changes (UML class diagram)}
    \label{fig:metamodel} 
\end{figure*}

\figurename~\ref{fig:metamodel} shows the proposed metamodel for replication changes using a UML class diagram. As can be seen, the classes at the top model the necessary context information for understanding replication changes, whereas the classes at the bottom model the structure and properties of the changes themselves. The \emph{enumerations} on the right of the diagram correspond to the categorical domains of some class attributes. %

With respect to the replication context, two types of empirical studies are considered, \emph{original studies} and \emph{replications}. Both of them are identified by an \emph{acronym}, take place in a given \emph{site} at a given \emph{date}, and usually have a \emph{report} that can be accessed using a URL. %
In the case of original studies, a specification of their \emph{goal}---that should include at least the \emph{cause} and \emph{effect} constructs and the \emph{population} under study, for example using a well-known template such as \GQM \cite{Basili1994}---and their \emph{description} are also included, since they are supposed to be shared by its replications, thus providing a context which is needed for understanding related replications and their changes. %
For replications, their type (\emph{internal}, \emph{external}), their purposes (\emph{confirm results}, \emph{generalize results}, or \emph{overcome limitations}), and their changes are recorded together with their \emph{base} study, i.e. the study they replicate, which can be an original study or a previous replication, as modeled by the abstract superclass \emph{EmpiricalStudy}.


Regarding replication changes, they are identified by a descriptive name and must describe narratively what changes from the base study (\emph{base\_situation}) to the replication (\emph{replication\_situation}). The \emph{purpose} of the changes must also be recorded, as well as their \emph{impacts} (if any) on experimental validity (modeled by the \emph{ChangeImpact} class), following the validity taxonomy described by \citeauthor{wohlin:experimentation} in \cite{wohlin:experimentation} and modeled by the \emph{Validity} enumeration. %
For each impact of a change, its \emph{rationale}, i.e. why the change affects a given validity type, must be recorded together with its effect, using a 7-point linear scale modeled by the \emph{Effect} enumeration. %
%
Note that associating a value in an linear scale from \emph{substantially} (-3), \emph{moderately} (-2), and \emph{slightly} (-1) decreases to their positive counterparts, including a \emph{does not affect} (0) central point, allows to have an idea on how the changes in the different experiments in a family increase or decrease original study's experimental validity and it can be easily visualized graphically, as it will be shown in Section \ref{sec:caesar}. 

\subsubsection{Change Dimensions} \label{sec:dimensions}

For classifying the different types of changes in replications, we have followed and expanded the \emph{dimensions} for experimental configuration proposed by \citeauthor{gomez2014understanding} in \cite{gomez2014understanding}, resulting in the classification described below. %
\begin{description}

\item[Operationalization change]
In a controlled experiment, the \emph{cause} construct is operationalized as one or more \emph{treatments}, whereas the \emph{effect} construct is operationalized as one or more \emph{metrics}, which are measured following \emph{measurement procedures}. Any change related to any of the above items (modeled as the \emph{OperationalizationElement} enumeration), e.g. changing the duration of a treatment, %
should be an instance of the \emph{OperationalizationChange} class. \\ 

\item[Population change] 
This class encompasses any change related to the \emph{experimental subjects} or \emph{objects}, e.g. changing the experience level or the average age of the experimental subjects with respect to those in the base study. \\
    
\item[Protocol change] 
As defined by \citeauthor{Juristo2012} in \cite{Juristo2012}, the experimental protocol is the setup of the \emph{experimental design}, \emph{experimental material}, 
\emph{experimental guides}, 
\emph{measuring instruments}, 
and \emph{data analysis techniques}, which are the elements modeled by the \emph{ProtocolElement} enumeration in our metamodel. Any change affecting one of those elements, such as changing the tasks that subjects have to perform or using Bayesian instead of frequentist statistics, should be an instance of the \emph{ProtocolChange} class. \\

\item[Experimenter change] These are changes related to the experimenters and their roles in the replication when compared to the base study. As modeled by the \emph{ExperimenterRole} enumeration, the roles considered are \emph{designer}, \emph{trainer}, \emph{monitor}, \emph{measurer} and \emph{analyst}. \\

\item[Context change] This class, not present in \cite{gomez2014understanding}, models any change related to the context in which the replication is carried out compared to the context in which the base study took place. For example, in human-oriented experiments using students as subjects, changing the time of year when the study is conducted from before final exams to after final exams is a change of this kind. In technology-oriented experiments, moving from running the software on a real machine to running it on a virtual machine would also be a context change.


\end{description}

\subsection{Templates for Replications Changes} \label{sec:plantilla}


Templates have been successfully applied in \CS, \SE, and related areas. For example, the 
well-known \GQM template proposed in \citeyear{Basili1994} by \citeauthor{Basili1994} \cite{Basili1994}, the \citeauthor{38631e0608b54d4299d5707f3a78debf}'s proposal for the specification of the problem under study in \DSR, and many others, including templates for requirements \cite{duran1999requirements,duran2002supporting}, process performance indicators \cite{del2012defining,del2016using}, or metamorphic testing relations \cite{segura2017template}. 



\begin{figure}[t]
  \footnotesize
  \sffamily 
	\centering  
	\renewcommand{\arraystretch}{1.25}
  
	\begin{tabularx}{0.9\textwidth}{|l|X|}
	\hline

	\textbf{Replication} & 
	\makecell*[lt]{\ph{repl. acronym} (\ph{report URL})\\[0.5mm]
	\{ Internal $\mid$ External \} replication based on \ph{base acronym}\\[0.5mm]
  \{ original study $\mid$ previous replication \}\\[1mm]} 
	\\ \hline

	\makecell*[lt]{\textbf{Original Study}} &
	\makecell*[lt]{\textbf{Goal}: \{ \ph{GQM} $\mid$ \ph{goal - cause, effect, population} \} \\[0.5mm] \textbf{Description}: \ph{description}\\[1mm]} 
	\\ \hline

	\makecell*[lt]{\textbf{Site and Date}} &
	\makecell*[lt]{The base experiment was carried out at \ph{site} in \ph{date} \\[0.5mm]
	This replication was carried out at \ph{site} in \ph{date}} 
	\\ \hline

  \textbf{Purposes} &
	\makecell*[lt]{ \{ $\bullet$~Confirm results\\$\:\mid$\, $\bullet$~Generalize results\\$\:\mid$\, $\bullet$~Overcome limitations of previous studies \}+\\[1mm]} 
	\\ \hline

	\textbf{Comments} &
	\makecell*[lt]{[\ph{comments}]}
	\\ \hline

	\end{tabularx}
  \vspace{0.5em}
	\caption{Template for replication context information}   
	\label{fig:context_template}
  
\end{figure}



\begin{figure}[t]
  \footnotesize
  \sffamily 
	\centering  
	\renewcommand{\arraystretch}{1.25}
  
 	\begin{tabularx}{0.9\textwidth}{|l|X|}
	\hline

	\textbf{Change \#}\ph{i} & 
	\makecell*[lt]{\ph{name} (\ph{repl. acronym})\\[1mm]} 
	\\ \hline

	\textbf{Description} & 
	\makecell*[lt]{Originally, \ph{situation in base experiment}.\\[0.5mm] 
	In this replication, \ph{situation in replication}.\\[0.5mm] 
  With the purpose of \ph{change purpose} \\[1mm]} 
	\\ \hline

	\makecell*[lt]{\textbf{Dimension}} &
	\makecell*[lt]{
	\{ Operationalization, specifically, the \\[0.5mm]
	\hspace{1em} \{ treatments $\mid$ metrics $\mid$ measurement procedures \} \\[0.5mm] 
	$\mid$ Population, specifically, the \ph{population property} \\[0.5mm] 
	$\mid$ Protocol, specifically, the \\[0.5mm]
	\hspace{1em} \{ experimental design $\mid$ experimental material \\[0.5mm]  
	\hspace{1em} $\mid$ experimental guides $\mid$ measuring instruments \\[0.5mm]  
	\hspace{1em} $\mid$ data analysis techniques \}\\[0.5mm] 
	$\mid$ Experimenter, specifically, the role of  \\[0.5mm]  
	\hspace{1em} \{ designer $\mid$ analyst $\mid$ trainer $\mid$ monitor $\mid$ measurer \}\\[0.5mm] 
	$\mid$ Context, specifically, the \ph{context variable} \} \\[1mm] 
	} \\ \hline

  \makecell*[lt]{\textbf{Effects on}\\\textbf{Validity}} & 
	\makecell*[lt]{ \{ $\bullet$ This change\\
  \hspace{1em} \{\ \{ substantially (3) $\mid$ moderately (2) $\mid$ slightly (1) \}\\ 
  \hspace{1.85em} \{ increases ($+$) $\mid$ decreases ($-$) \} \\
  \hspace{1em} $\mid$ does not affect (0) \}\\
  \hspace{1em} \ph{validity type} because \ph{rationale}\\
  \}+ \\[1mm] 
	} \\ \hline

	\textbf{Comments} &
	\makecell*[lt]{[\ph{comments}]}
	\\ \hline

	\end{tabularx}
  \vspace{0.5em}
	\caption{Template for replication changes (one for each change in the replication)}   
	\label{fig:change_template} 
\end{figure}

Templates help visualizing information in a standard form, which can be easily adopted by practitioners, especially by novice researchers. In order to improve its usability, we have augmented templates with \emph{linguistic patterns} (L-patterns) when possible, in a similar manner than in %
\cite{duran1999requirements,duran2002supporting,del2012defining,del2016using}. %
L-patterns are pre--written, parametrized sentences that can be used for filling in some template fields in a easier, standard way. %
Driven by the metamodel in \figurename~\ref{fig:metamodel}, our proposal includes two templates, one for the replication context shown in \figurename~\ref{fig:context_template}, and another for replication changes shown in \figurename~\ref{fig:change_template}. %

In the notation used in the templates in \figurename{}s~\ref{fig:context_template} and \ref{fig:change_template}, placeholders are depicted between angle brackets \hbox{(\ph{\ldots\!})}, single options are enclosed by curly brackets and separated by vertical bars \hbox{(\{\ldots $\vert$\ldots \})}, multiple options are indicated by a plus symbol after the closing bracket \hbox{(\{\ldots $\vert$\ldots \}+)}, and elements in square brackets \hbox{([\ldots\!])} are considered as optional.

\subsection{Proof-of-concept Software Tool: CÆSAR} \label{sec:caesar}

\begin{figure}[b]
	\centering
	\includegraphics[scale=0.45,trim={0cm 1.75cm 0cm 2.65cm},clip]
  {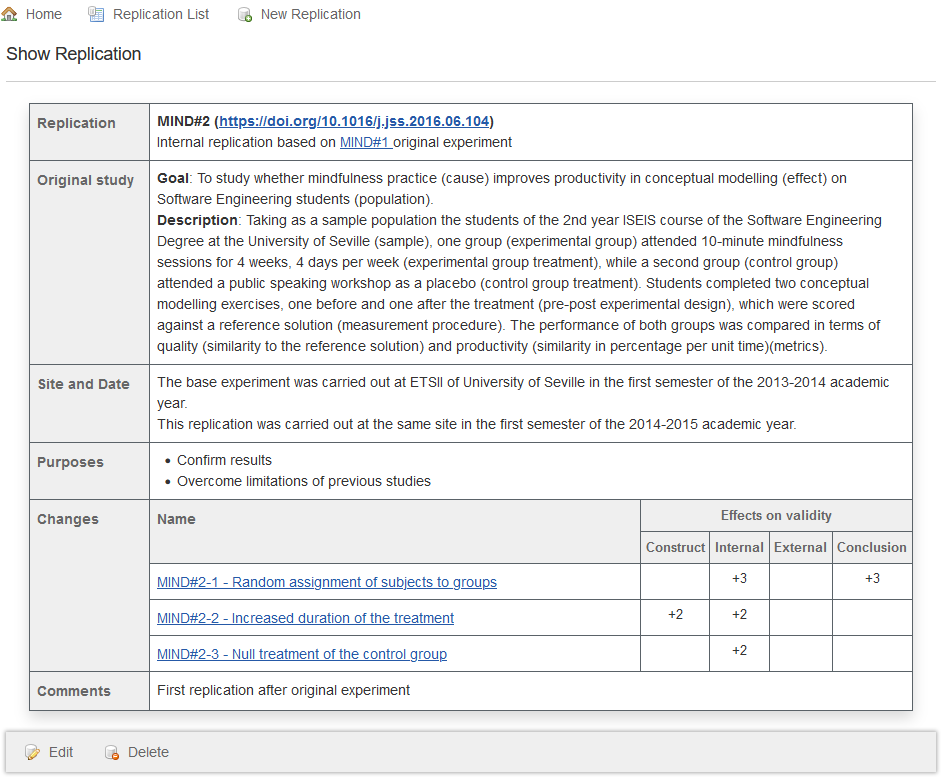}
  \vspace{0.5em}
	\caption{Example of context template in \caesar for the replication reported in \cite{bernardez-jss-2016}}
	\label{fig:mind2_template}
  \vspace{0.5em}
  \includegraphics[scale=0.45,trim={0cm 1.75cm 0cm 2.75cm},clip]
  {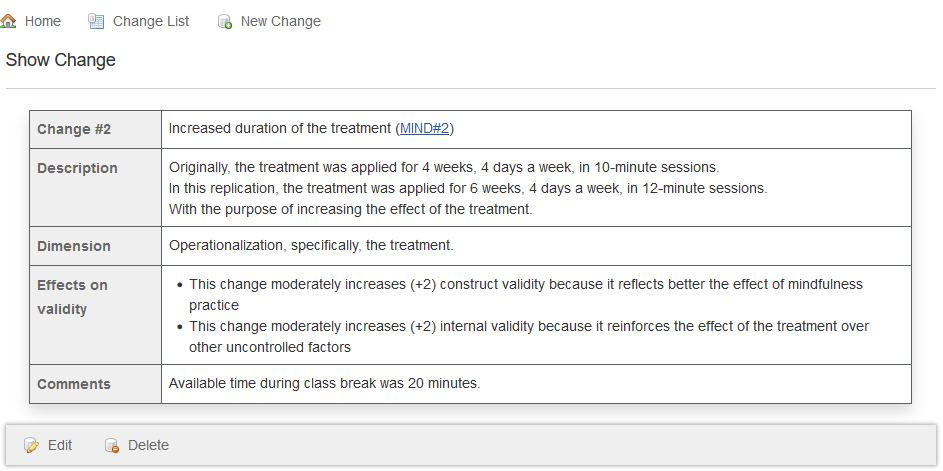}
  \vspace{0.5em}
	\caption{Example of change template in \caesar for one of the changes in \cite{bernardez-jss-2016}}
	\label{fig:mind2_2_template}
\end{figure}

Applying a model-driven development approach, we have built a proof-of-concept tool called \caesar (ChAngE SpecificAtion for Replications) using the Grails framework (\url{https://grails.org/}) and the metamodel described in Section \ref{sec:metamodel}. 
\caesar is deployed for demonstration at \caesarurl, providing a simple interface for manipulating instances of the entities in the metamodel and displaying their information using the templates described in the previous section, as shown in \figurename{}s~\ref{fig:mind2_template} and \ref{fig:mind2_2_template}. %

As can be seen in \figurename~\ref{fig:mind2_template}, the context template for replications have been augmented with information corresponding to the effects of their changes---expressed in a 7-point linear scale from -3 to +3 as commented in Section~\ref{sec:metamodel}---on the four types of experimental validity proposed in \cite{wohlin:experimentation}, which can be used by experimenters when reporting their replications according to their preferences.

\begin{figure}[b]
	\centering
	\includegraphics[scale=0.45]{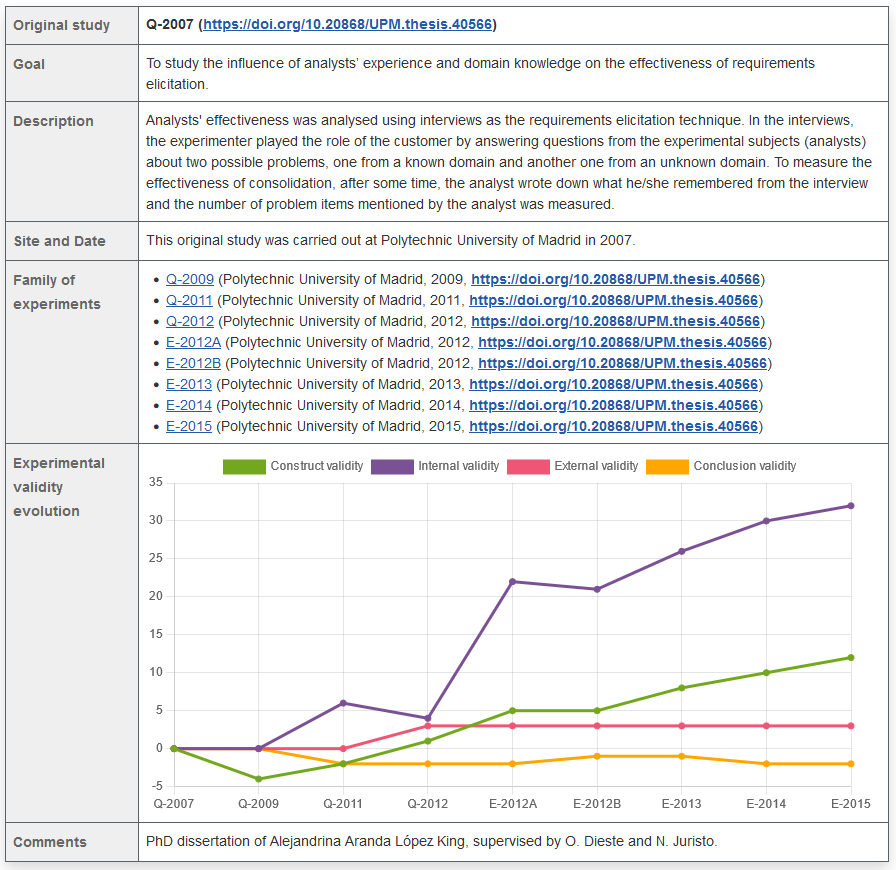}
  \vspace{0.5em}
	\caption{Validity evolution for the family of experiments described in \cite{aranda2016estudio}}
	\label{fig:caesar_graph_q_2007}
\end{figure}

\begin{figure}[t]
	\centering
	\includegraphics[scale=0.45]{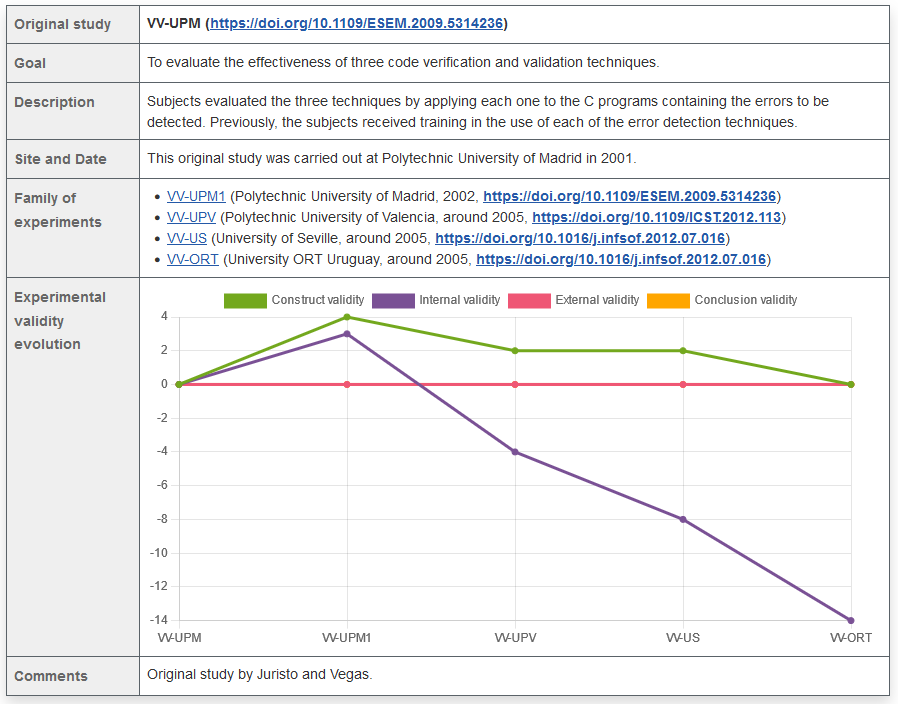}
  \vspace{0.5em}    
	\caption{Validity evolution for the family of experiments described in \cite{juristo2003functional,juristo2012comparing,juristo2013process}}
	\label{fig:caesar_graph_vv_upm}
\end{figure}

Apart from validating the metamodel displayed in \figurename~\ref{fig:metamodel} and adding computed information to the templates, the main added value of building this tool is the possibility of visualizing the evolution of the experimental validity along a family of experiments. For that purpose, a template for original studies has also been developed into the tool in which all the studies in an associated family of experiments are queried and the effects of their changes are displayed accumulatively from the original study---for which all types of validity are assumed to start at zero---as a line graph for the four aforementioned types of experimental validity, as shown in \figurename{}s~\ref{fig:caesar_graph_q_2007} and \ref{fig:caesar_graph_vv_upm}. %

This kind of visualization provides a valuable information about a family of experiments that is not usually reported and that can drive changes in new replications. For example, note how in the case of the family of experiments described in \cite{aranda2016estudio} and displayed in \figurename~\ref{fig:caesar_graph_q_2007}, three types of experimental validity increase along time---especially internal validity---whereas in the family described in \cite{juristo2003functional,juristo2012comparing,juristo2013process} and displayed in \figurename~\ref{fig:caesar_graph_vv_upm}, internal validity substantially decrease, mainly because of the lack of resources 
in follow-up replications, as commented later in Section~\ref{sec:hoe-case}. %
In the latter case, the graph clearly indicates that experimenters carrying out a new replication should try to increase internal validity whenever possible.

\caesar has been successfully used in the validation of our proposal, which is discussed in the next section.


\section{Multiple Case Study for Artifact Evaluation} \label{sec:case_study}

As commented in \cite{wohlin:experimentation}, experiments in \CS may be classified as \emph{human-oriented} or \emph{technology-oriented} depending mainly on the nature of experimental subjects. In order to evaluate the suitability of our artifact, we conducted a multiple case study involving both human and technology-oriented families of experiments in \CS. %

Since we had the opportunity to meet experimenters from a different discipline, \Agrobiology, we decided to include some families of experiments from that discipline in the multiple case study with a twofold purpose. %
On one hand, to check whether our artifact could also be applied to a completely different type of experiments, using plants as subjects, which somehow have some common characteristics not only to human-oriented experiments, but also to technology-oriented ones. On the other hand, to identify reported information that could be incorporated into the templates and thus improve them.

\begin{figure*}[t]
  \centering
  \includegraphics[scale=0.60,trim={0.75cm 9.75cm 10.00cm 1.00cm},clip]
  {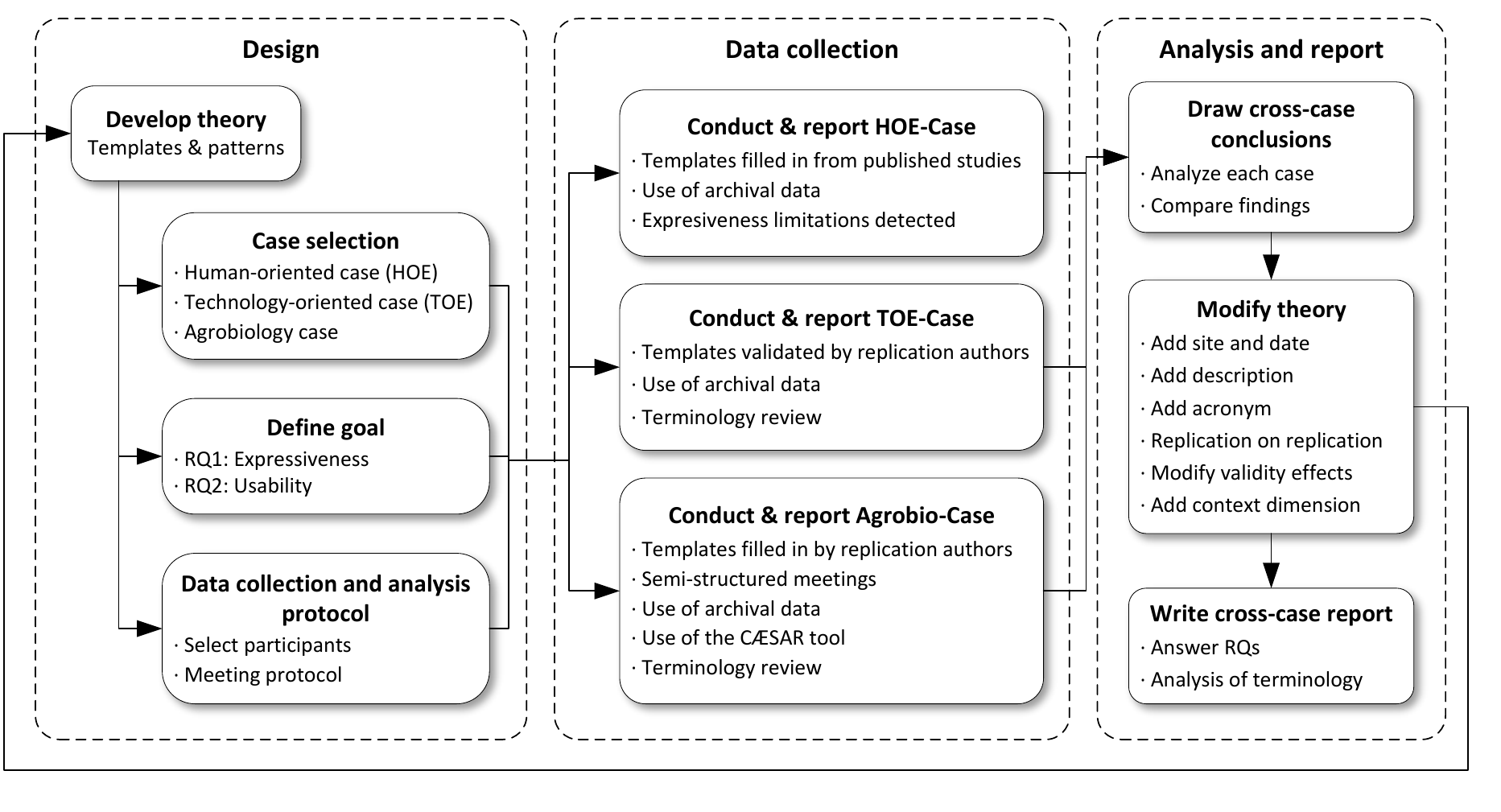}
  \caption{Phases of the multiple case study for artifact evaluation} 
  \label{fig:Multiple-CaseStudy} 
\end{figure*}

The research method followed for such evaluation was based on the case study research process proposed by \citeauthor{runeson2012case} in \cite{runeson2009guidelines,runeson2012case}. %
An overview of the phases of the multiple case study, which are detailed in the next sections, is depicted in \figurename~\ref{fig:Multiple-CaseStudy}.

\subsection{Design and Data Collection} \label{sec:case_study_design}

In the design phase, those families of experiments to evaluate our proposal were selected applying the criteria of having a relevant number of replications and changes in order to cover as many different options in the proposed templates as possible and to answer the following research questions (RQs): 
 
\begin{description}

\item[\textnormal{\RQ{1} (\Expressiveness)}]
Can the proposed templates be used to specify reported replication changes properly? Do they need to be augmented to include more reported information? Is the usually reported information sufficient to specify all the information included in the templates? Are they suitable for disciplines other than \CS?\\
 
\item[\textnormal{\RQ{2} (Usability)}]
Do researchers find the proposed templates useful? Do they find them easy to use? Is the terminology used understood by researchers outside \CS community?
        
\end{description} %
%
%

For each case study, a brief description of the selected families, the followed protocol, and the collected data 
is presented below. %
Note that the collected data, i.e. the specifications of all the selected replication studies and their changes using a \LaTeX\ version of the proposed templates, is publicly available at Zenodo \cite{cruzzenodo2021}, although the meeting minutes are not included for the sake of the privacy of the participant researchers.

\subsubsection{Human-Oriented Case\ (\case{\HOE--Case})} \label{sec:hoe-case}

 
For this first case study, we selected three families of human-oriented experiments dealing with (i) the effect of mindfulness on conceptual modeling performance (\case{\Mind}) \cite{bernardez2014controlled,bernardez-jss-2016,bernardez2020effects}, (ii) requirements elicitation (\case{\Req}) \cite{aranda2016estudio}, and (iii) code testing techniques (\case{\Test}) \cite{juristo2003functional,juristo2012comparing,juristo2013process}, due to our familiarity with these topics. 
This case study was designed as a self--evaluation case study, so all the replications and changes were specified by ourselves after a close reading of the corresponding reports. %
All the limitations and issues found were registered for further analysis and potential artifact improvement.

\subsubsection{Technology-Oriented Case (\case{\TOE--Case})} \label{sec:toe-case}
 
To evaluate our proposal with other types of \CS experiments where the subjects are not human beings, %
two families of experiments on automated software testing (\case{\Testing}) \cite{parejo2016multi}, and software product line testing (\case{\SPL}) \cite{sanchez2014comparison} were also selected at suggestion of the researchers who accepted to participate and to whom we had direct access. %
In this case study, the selected replication changes were also specified by ourselves, but then several meetings were held with the researchers who carried out the experiments to validate our specifications and obtain feedback from them, recording the valuable information. %

\subsubsection{\Agrobiology (\case{\Agrobio--Case})}


In order to evaluate our proposal in other areas of knowledge, we selected four families of experiments belonging to \Agrobiology dealing with soil decontamination (\case{\Soil}) \cite{Sheila2016,marquez2018,Guijarro2019TFG}, harvesting systems (\case{Harvest}) \cite{plasquy2021florido},  extraction of olive oil components (\case{\Olive}) \cite{garcia2016extraction}, and influence of diet on cholesterol accumulation (\case{\Diet})	\cite{pacheco2008meal}. These families %
were selected at the suggestion of the \Agrobiology researchers who accepted to participate and to whom we had direct access. %

Several meetings were held where we followed a protocol consisting of the following steps: %
(i) we explained the templates to ensure that the researchers understood how to use them; %
(ii) we asked each researcher to select one of her experiments with at least one replication and, if possible, already published; %
(iii) for the selected experiment, we asked each researcher to fill in the corresponding templates for all the experiment replications and their associated changes, providing us with as much feedback as possible; %
and (iv) we asked the researchers whether they had found any limitations, usability problems, or unknown terminology using the templates, recording all the details.
%
By the time we conducted this case study, an earlier version of the \caesar tool was already available, and some researchers agreed to use it for reporting their replications, providing very valuable feedback. 

\subsection{Analysis and Report} \label{sec:Joint}

In this section, the results of the three case studies, which are summarized in \tablename~\ref{tab:summary}, are analyzed and reported.

%



\begin{sidewaystable}[tbp]
  \caption{Summary of the multiple case study outcomes}
  \vspace{0.5em}
  \label{tab:summary}
  \centering
	\footnotesize
  \scalebox{0.95}{%
  \begin{tabularx}{\textwidth}{ccccc>{\hsize=.90\hsize}X>{\hsize=1.10\hsize}X}
    \toprule
   
  	\makecell*[cc]{Case\\study}     & 
    \makecell*[cc]{Family\\of Exp.} & 
    \#Repl & 
    \#Chng & 
	  \makecell*[cc]{Originally\\Reported as} & 
    \makecell*[cc]{Artifact evolution}      & 
    \makecell*[cc]{Additional findings} \\
	 
    \midrule
 	 	\addlinespace
    
 	  \case{\HOE} 
    
		& \case{\Mind} & 2 & 4 & Tabular 
    & Some replications were based on replications, not only in original studies. This results in the new abstract superclass \emph{EmpiricalStudy} and the application of the \emph{Composite} design pattern. Site, date, and acronym attributes are added to the new class. 
    & Except for the replications based on replications, 
 		all changes (4 out of 4, 100\%) could be specified with the early version of the metamodel and templates. \\
 		
 		& \case{\Req}	& 8 & 33 & Tabular 
    & Some changes impact more than one validity type. This results in an evolution of the structure for modeling changes, impacts and effects. Rationale attribute is added to \emph{ChangeImpact} class. 
    &	11 out of 33 (33\%) changes could not be entirely specified with the early version of the metamodel and templates due to impacting more than one validity type. \\

 		& \case{\Code}	& 4 & 21 & Tabular  
    & The dimension of some changes could not be clearly identified, resulting in the new \emph{ContextChange} subclass.
    & 3 out of 21 (14\%) changes could not be completely specified for the same reason than in the \case{Req} family and because of the lack of the \emph{context} change dimension. \\
 		
  	 	
    \makecell*[cr]{\textbf{Total}} & \textbf{3} &  \textbf{14} & \textbf{58} & & & 
    In general, change purpose is seldom reported. Impacts on validity and modified dimensions are not always straightforward to identify.\\

 	 	\addlinespace
		\midrule
 	 	\addlinespace		

    \case{\TOE}
 		
		& \makecell*[tc]{\case{\Testing}\\[0.1cm]\case{SPL}}	
    & \makecell*[tc]{3\\[0.1cm]1}
    & \makecell*[tc]{3\\[0.1cm]2} 
    & \makecell*[tc]{Narrative\\[0.1cm]Narrative} 
    & No relevant evolution of the artifact as a consequence of this case study.
    & This type of experiments could be properly specified using an evolved version of the metamodel and templates, although the specification of change dimensions was not always obvious due to differences with \HOE studies.\\[-0.25cm]
 		
  	 	
    \makecell*[cr]{\textbf{Total}} & \textbf{2} &  \textbf{4} & \textbf{5} & & &  	
		\\

 	 	\addlinespace
		\midrule
 	 	\addlinespace		

    \case{\Agrobio}
 		
		& \makecell*[tc]{%
        \case{\Soil}\\[0.1cm]
        \case{\Harvest}\\[0.1cm]
        \case{\Olive}\\[0.1cm]
        \case{\Diet}
      }	
        
    & \makecell*[tc]{%
        2\\[0.1cm]
        1\\[0.1cm]
        1\\[0.1cm]
        1
      }  
      
    & \makecell*[tc]{%
        16\\[0.1cm]
        1\\[0.1cm]
        11\\[0.1cm]
        1
      }

    & \makecell*[tc]{
        Narrative\\[0.1cm]
        Narrative\\[0.1cm]
        Tabular\\[0.1cm]
        Narrative
      } 
    
		& A clear need of contextual information related to the original study of a family of experiments is detected. This results in new attributes in the \emph{OriginalStudy} superclass and it is the motivation for the quantification of effects on validity and its visualization in the \caesar tool in further iterations.
    
    & Without contextual information, replications are almost impossible to understand. Specification of change dimensions was not obvious. Some differences in experimental terminology were detected. Positive feedback with respect to the the \caesar tool. \\[-0.25cm]
 	
  	 	
    \makecell*[cr]{\textbf{Total}} & \textbf{4} &  \textbf{5} & \textbf{29} & & & \\ 	
	   
 	 	\addlinespace
		\midrule

    \makecell*[cr]{\textbf{Total}} & \textbf{9} &  \textbf{23} & \textbf{92} & & & \\

    \bottomrule
    

	\end{tabularx}
  } 
\end{sidewaystable}

\subsubsection{\case{\HOE--Case} analysis}

\begin{figure}[b!]%
  \centering
  \subfigure[Early version]{
    \includegraphics[scale=0.625,trim={2.75cm 21.60cm 13.75cm 2.35cm},clip]  
    {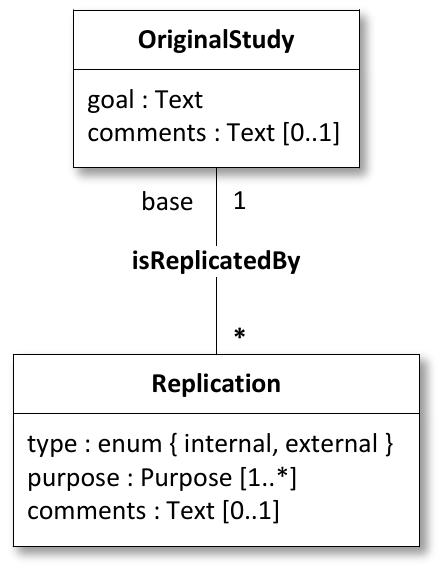}
    \label{fig:metamodel_evolution_01_a}
  }\qquad\qquad
  \subfigure[Evolved version]{
    \includegraphics[scale=0.625,trim={8.5cm 21.60cm 4.75cm 2.20cm},clip]  
    {figures/metamodel_evolution.pdf}
    \label{fig:metamodel_evolution_01_b}
  }
  \caption{Metamodel evolution for modeling replications based on replications}
  \label{fig:metamodel_evolution_01}
  \vspace{0.5em}  
  \subfigure[Early version]{
    \includegraphics[scale=0.62,trim={1.415cm 16.60cm 15.10cm 9.60cm},clip]  
    {figures/metamodel_evolution.pdf}
    \label{fig:metamodel_evolution_02_a}
   }\quad
  \subfigure[Evolved version]{
    \includegraphics[scale=0.62,trim={7.10cm 16.60cm 1.75cm 9.25cm},clip]  
    {figures/metamodel_evolution.pdf}
    \label{fig:metamodel_evolution_02_b}
  }
  \caption{Metamodel evolution for modeling more than one validity effect per change}
  \label{fig:metamodel_evolution_02}
\end{figure}

When this first case study was conducted, the metamodel of our artifact was in a early stage. This is the reason why we designed it as a self--evaluation case study and that is why most of the evolution of the metamodel---and therefore of the templates and the \caesar tool---is based on the findings obtained during its conduction. As a matter of fact, only 76\% of the changes in the selected families of experiments could be specified with the initial version of the metamodel, but 100\% with the evolved version. %
One of the main evolutive changes in the early metamodel was motivated by the fact that some of the replications in the \case{Mind} family were based not on an original study but on a previous replication. Since replications in the initial metamodel were associated to original studies only, we had to include the abstract class \emph{EmpiricalStudy} to allow replications based on replications, in a similar way to the \emph{Composite} design pattern \cite{DesignPatterns}. %
As shown in \figurename~\ref{fig:metamodel_evolution_01}, in this evolutionary change we also introduced some missing information like \emph{date}, \emph{site}, and \emph{report} and an \emph{acronym} as a public unique identifier to facilitate the referencing of empirical studies and their traceability.

On the other hand, during the specification of the replication changes in this case study, %
we observed that some of them affected more than one type of experimental validity, so we evolved the metamodel accordingly, as shown in \figurename~\ref{fig:metamodel_evolution_02}. %
Sometimes, the impact of a change on a specific type of experimental validity was not obvious, so we also decided to add the \emph{rationale} attribute to the new \emph{ChangeImpact} class, letting the experimenters register why they thought a given change increased or decreased a given validity type. %
In a subsequent iteration, we considered that this effect could be subjectively quantified using a 7-point linear modeled by the \emph{Effect} value-based enumeration and the corresponding \emph{effect} attribute in the \emph{ChangeImpact} class. %
As previously commented in Section~\ref{sec:metamodel}, this enhancement of the metamodel allows to visualize the evolution of the experimental validity across a family of experiments (see \figurename{}s~\ref{fig:caesar_graph_q_2007} and \ref{fig:caesar_graph_vv_upm}), supporting decisions in further replications.
 

%
It is worth mentioning that in most of the reported replications the purposes of the changes were rarely stated, making it difficult to identify their effects on validity and the modified dimensions. 

\subsubsection{\case{\TOE--Case} analysis} \label{sec:CompCaseAnalysis}

The main goal of this case study was to confirm the results of the previous one and to consolidate the evolutive changes on the artifact. %
Using the evolved templates, we were able to specify all the replications changes, although it took us some time to agree on the dimensions modified by some changes.

Overall, considering that most of the limitations of the initial artifact had been identified in the first case study, the participating researchers found our approach satisfactory, providing very positive feedback.

\subsubsection{\case{\Agrobio--Case} analysis}

\begin{figure}[htb]
	\centering
  \includegraphics[scale=0.45]{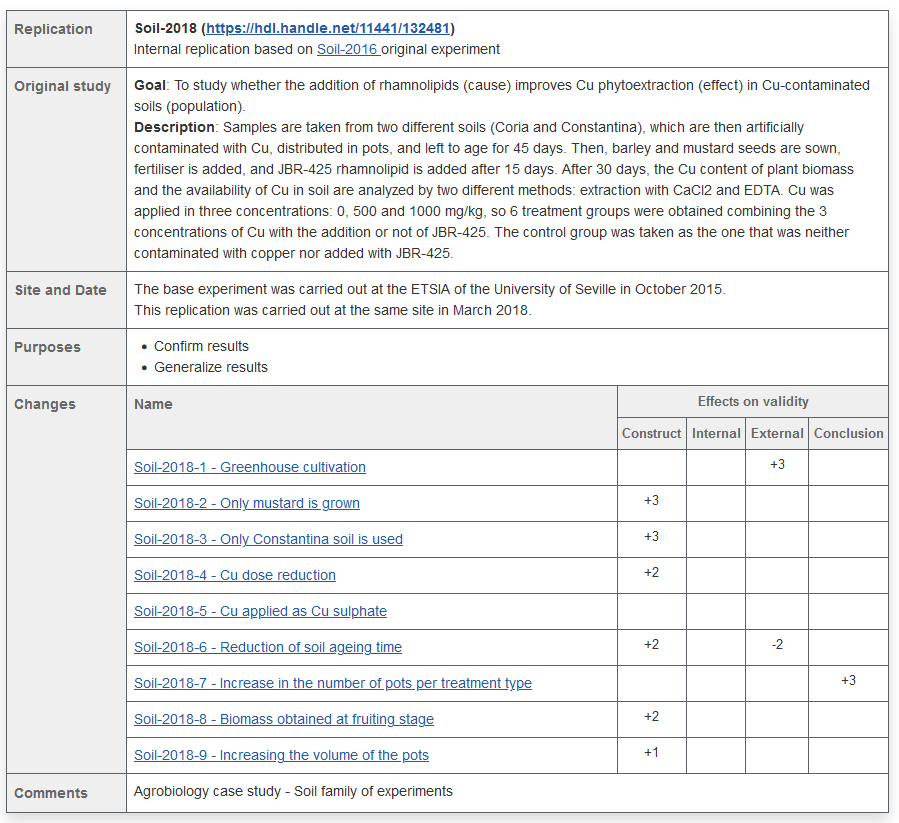}
  \vspace{0.5em}
	\caption{Specification of the \case{Soil--2018} replication \cite{marquez2018} in the \caesar tool}
	\label{fig:caesar-soil2018}
\end{figure}

After validating our approach for the specification of changes in \CS experiments, the main goal of the third case study was to test it in a different research area with users other than ourselves. %
One of the first issues identified during this case study was the need of a context for replication changes to be understood due to their high specificity. Clearly,  that context had to be provided by the original study since it was shared by all its replications. %
As shown in \figurename~\ref{fig:metamodel_evolution_01}, in the early version of the metamodel, only the goal of the original study was registered whereas in the evolved metamodel, a description is added to provide such a context. See the description of the original study of the \case{Soil--2018} replication in \figurename~\ref{fig:caesar-soil2018} for a clear example.

In families of experiments in \Agrobiology, it is a common practice to change the growing medium of the plants from Petri dishes to culture chambers, to greenhouses, and finally, to natural soils. It is also common to repeat the same experiment in different seasons to confirm results. %
These kinds of change, that we were not able to classify in any of the change dimensions by \citeauthor{gomez2014understanding} \cite{gomez2014understanding}, were the reason to expand their proposal with the \emph{context}\footnote{The name was decided because this kind of change usually affects \emph{context variables} in experimental design. An alternative name, \emph{environment} dimension, was also discussed among us but was finally discarded in favor of \emph{context}.} dimension. %
Including this new change dimension, we were also able to classify some changes in the \case{\HOE--Case} in which the experiments were held with students at different moments of the academic course, e.g., at the beginning of the course, during examination period, and that we had not been able to completely specify before. %

Regarding the user experience of the participating researchers, they found the templates and L--patterns very useful for reporting their replications, helping them to focus on the essential information. %
Those who agreed to use \caesar, despite being a proof--of--concept tool, found it very useful for generating reports of their experiments and showed a great interest to use it in the future.
Nevertheless, some differences in the terminology used in the templates were detected. %
Particularly, the concept of \emph{change dimension} was foreign to them, something that it seemed reasonable considering that it is a recent concept proposed by an \CS researcher \cite{gomez2014understanding}. %
Another foreign concepts to them were those of \emph{experimental validity} and \emph{validity threats}, which was very surprising to us considering the importance they are given in our discipline. After several meetings, we found out that they handle a list of usual experimental risks with their corresponding actions to mitigate them, but they use the term \emph{avoiding experimental errors} instead of \emph{minimizing validity threats} as we do in \CS. %
On the other hand, they preferred the term \emph{repetition} to \emph{replication}. 
\subsection{Threats to Validity} \label{sec:Validity}

\citeauthor{runeson2012case} \cite{runeson2012case} provide a detailed description of threats to the validity of case studies. In this section, such threats and the actions performed to mitigate them in our multiple case study are described. 

\subsubsection{Construct validity}

This validity is concerned with the degree to which the operationalization actually reflects what is to be the investigated. According to \cite{yin2003design}, using multiple sources of evidence during data collection increases construct validity, %
as it has been our case. %
%
On the other hand, it is also recommended in \cite{yin2003design} that a draft of the case study report be reviewed by key informants. %
In our case, the participant researches have not only followed the evolution of the artifact across the whole evaluation process, but they have also reviewed a draft report of our multiple case study.


    
\subsubsection{Internal validity}

This validity is concerned when causal relations are examined. %
%
The main hypothesis in our study is that the use of the proposed artifact is useful for the specification and reporting of replications, not only in \CS, but also in other areas of knowledge. 
The only potential threat to the internal validity is that the differences in the used terminology could affect the usage of the artifact by participant researchers, especially for those from research areas other than \CS. %
To mitigate this threat, we explained the use of the templates and assisted the \Agrobiology researchers during the \case{\Agrobio--Case}, apart from investigating the differences and similarities between their terminology and that used in \CS, in order to improve our communication with them.


\subsubsection{External validity}

This validity is concerned with the generalization of results in cases that have common characteristics. %
The fact that the templates were firstly used by ourselves 
could limit the generalization of the findings. In order to overcome this threat, we invited other researchers from our own research area and from other research areas to participate in the study, thus reducing the potential bias. Nevertheless, further research is needed to validate the artifact in more replications, belonging to the \CS discipline or to other different disciplines. %
In this sense, the \caesar tool is available to the scientific community, so other researchers can report their replication changes using it and provide feedback for further improvements. 

\subsubsection{Reliability}

This aspect is concerned with obtaining similar results when the study is carried out by other researchers. %
Since the specification of all replications with their corresponding changes 
is available at Zenodo \cite{cruzzenodo2021}, the study can be repeated by other researchers which can then compare their results with ours.




\section{Related work} \label{sec:trabajos}

Regarding replications in \CS, %
several literature reviews show the relevance of the topic \cite{bezerra2015replication,cruz2019replication,da2014replication,de2015investigations}. %
These reviews are systematic mapping studies dealing with general aspects such as types of replications, conceptual frameworks, or addressed research topics. 
However, to the best of our knowledge, there is a lack of literature reviews on more specific aspects of replications such as the reporting of changes. %
Despite of the recommendations by \citeauthor{carver2010towards} \cite{carver2010towards}, \citeauthor{shepperd2018role} \cite{shepperd2018role}, \citeauthor{juristo2009using} \cite{juristo2009using,juristo2011role}, \citeauthor{gomez2014understanding} \cite{gomez2014understanding}, and \citeauthor{Baldassarre} \cite{Baldassarre} to report replication changes, no concrete proposals have been provided on how to specify them in \CS. %

The works that are most related to ours are 
\cite{1330459}, 
\cite{solari2013identifying}, and 
\cite{vegas2020mis}, who propose a tabular form to summarize the experiments that compose a family. %
In \cite{1330459}, replications are reported including their motivation, their changes (in  unstructured narrative text), the confirmation or non-confirmation of results in previous experiments, other characteristics such as subjects, tasks, and materials, and whether hypotheses or research questions changed from previous experiments. %
In \cite{solari2013identifying}, a tabular replication summary is used, %
including information on, among others, experimental design, laboratory package, material preparation, replication operation, data analysis, and experimenter evaluation. %
Finally, in \cite{vegas2020mis}, the specification of each experiment is summarized including elements such as factors, treatments, response variables, design, experimental objects, and participants.

After reviewing these proposals, we consider that all of them provide an in-depth overview of the configuration of each experiment in a family. %
However, to identify replication changes, table content corresponding to the different replications must be compared, whereas in our proposal, apart from being model-based and having tool support, each change is specified separately in a explicit manner, thus reducing tacit knowledge as recommended in \cite{kitchenham2008role}. %
%
%

%
%
In a recent survey closely related to the approach presented in this article, 137 articles published between 2013 and 2018 reporting at least one replication %
were analyzed \cite{cruz2019replication}. %
Most of the replication changes were defined in controlled experiments, 
confirming that this is the most frequently applied empirical strategy in the research area \cite{Guevara-Vega2021}. 
In general, most changes---also referred to as \emph{adjustments} or \emph{differences}---were described with respect to an original experiment using natural language or tabular forms, only about 25\%-30\% of the studies reported the purpose of the changes and their effect on experimental validity. %

\section{Conclusions and Future work} \label{sec:conclusions}

In the context of replications of empirical studies in \CS, and applying \DSR, we have develop and evaluated a composite artifact to systematically specify and report replication changes. %
We have developed the artifact following a model-based approach, generating (i) a metamodel formalizing all the relevant concepts related to replication changes; (ii) templates for reporting replication contexts and replication changes using the information in the metamodel together with L-patterns to facilitate their use; and (iii) a proof-of-concept tool based on the metamodel and supporting the management of the proposed templates. %
We have also evaluated the artifact by means of a multiple case study in which 92 replications changes corresponding to 23 replication studies in 9 families of experiments in the areas of \CS and \Agrobiology were specified by ourselves and by other researchers and subsequently analyzed. %
%
The evaluation revealed some initial limitations of our approach, but they were overcome after several improvement iterations. One of the most relevant improvements was the quantitative determination of the effect of changes on experimental validity, that allows the visualization of the evolution of the validity of a family of experiments thus supporting decision-making.

%
%
Our model-based proposal for the specification of replication changes seems to fit the needs not only of \CS, but also of other research areas such as \Agrobiology, obtaining very positive feedback from the participant researchers from both disciplines.

As future work, our aim in the middle term is to apply our templates not only to report changes of already conducted replications, but to use them during replication design, as a means of analyzing and documenting the purpose and effects of replication changes before they are performed. %
This design-oriented approach needs a more advanced version of the \caesar tool, including a virtual assistant, probably a \emph{chatbot}, that guides the researcher in the process and suggests, for example, potential effects of changes on experimental validity depending on the modified dimension or on other criteria that might be identified from experience. %
In the short term, we want to provide \CS researchers with a consolidated \LaTeX\ package for using the proposed templates in their articles when they need to report some replication changes, thus disseminating our artifact in the community. %

\section*{Acknowledgments}

We want to thank Carmen Florido, Aránzazu García, Rocío Abia, and Eddy Plasquy %
for their contribution to the multiple case study by specifying their experiments and for their kind willingness to provide us with all the information required.

This work has been partially supported by the European Commission (FEDER) and the Spanish Government under projects OPHELIA (RTI2018-101204-B-C22),
EKIPMENT-PLUS (P18-FR-2895), and MEMENTO (US-1381595).

\bibliographystyle{sn-basic}
\bibliography{Plan-bib}

\begin{thebibliography}{73}
\providecommand{\natexlab}[1]{#1}
\providecommand{\url}[1]{{#1}}
\providecommand{\urlprefix}{URL }
\providecommand{\doi}[1]{\url{https://doi.org/#1}}
\providecommand{\eprint}[2][]{\url{#2}}
 \bibcommenthead

\bibitem[{Albayrak and Carver(2014)}]{albayrak2014investigation}
Albayrak {\"O}, Carver JC (2014) Investigation of individual factors impacting
  the effectiveness of requirements inspections: a replicated experiment.
  Empirical Software Engineering 19(1):241--266.
  \doi{10.1007/s10664-012-9221-0}

\bibitem[{Almqvist(2006)}]{1330459}
Almqvist JPF (2006) Replication of controlled experiments in empirical software
  engineering - a survey. Master's thesis, master’s thesis, Department of
  Computer Science, Faculty of Science, Lund University, Sweden

\bibitem[{Aranda(2016)}]{aranda2016estudio}
Aranda A (2016) Empirical study of the influence of analyst experience and
  domain knowledge on the effectiveness of requirements education. PhD thesis,
  Polytechnic University of Madrid, \doi{10.20868/UPM.thesis.40566}

\bibitem[{Assar et~al(2016)Assar, Borg, and Pfahl}]{assar2016using}
Assar S, Borg M, Pfahl D (2016) Using text clustering to predict defect
  resolution time: a conceptual replication and an evaluation of prediction
  accuracy. Empirical Software Engineering 21(4):1437--1475.
  \doi{10.1007/s10664-015-9391-7}

\bibitem[{{Association for Computing Machinery}(2020)}]{ACM-2020}
{Association for Computing Machinery} (2020) Artifact review and badging.
  \url{https://www.acm.org/publications/policies/artifact-review-and-badging-current}

\bibitem[{Baldassarre et~al(2014)Baldassarre, Carver, Dieste, and
  Juristo}]{Baldassarre}
Baldassarre MT, Carver J, Dieste O, et~al (2014) Replication types: Towards a
  shared taxonomy. In: Proceedings of EASE'14, p 18:1–18:4,
  \doi{10.1145/2601248.2601299}

\bibitem[{Basili et~al(1994)Basili, Caldiera, and Rombach}]{Basili1994}
Basili VR, Caldiera G, Rombach HD (1994) Goal question metrics paradigm.
  Encyclopedia of Software Engineering pp 528--532

\bibitem[{Basili et~al(1999)Basili, Shull, and Lanubile}]{basili1999building}
Basili VR, Shull F, Lanubile F (1999) Building knowledge through families of
  experiments. IEEE Trans Softw Eng 25(4):456--473. \doi{10.1109/32.799939}

\bibitem[{Bern{\'a}rdez et~al(2014)Bern{\'a}rdez, Dur{\'a}n, Parejo, and
  Ruiz-Cortés}]{bernardez2014controlled}
Bern{\'a}rdez B, Dur{\'a}n A, Parejo JA, et~al (2014) A controlled experiment
  to evaluate the effects of mindfulness in software engineering. In:
  Proceedings of ESEM'14, pp 17--27, \doi{10.1145/2652524.2652539}

\bibitem[{Bern{\'a}rdez et~al(2020)Bern{\'a}rdez, Durán, Parejo, Juristo, and
  Ruiz-Cortes}]{bernardez2020effects}
Bern{\'a}rdez B, Durán AD, Parejo JA, et~al (2020) Effects of mindfulness on
  conceptual modeling performance: A series of experiments. IEEE Transactions
  on Software Engineering 48(2):432--452. \doi{10.1109/TSE.2020.2991699}

\bibitem[{Bernárdez et~al(2018)Bernárdez, Durán, Parejo, and
  Ruiz-–Cortés}]{bernardez-jss-2016}
Bernárdez B, Durán A, Parejo JA, et~al (2018) An experimental replication on
  the effect of the practice of mindfulness in conceptual modeling performance.
  Journal of Systems and Software 136:153--172. \doi{10.1016/j.jss.2016.06.104}

\bibitem[{Bezerra et~al(2015)Bezerra, da~Silva, Santana, Magalhaes, and
  Santos}]{bezerra2015replication}
Bezerra RM, da~Silva FQ, Santana AM, et~al (2015) Replication of empirical
  studies in software engineering: An update of a systematic mapping study. In:
  Proceedings of ESEM'15, IEEE, pp 1--4, \doi{10.1109/ESEM.2015.7321213}

\bibitem[{Brooks et~al(1996)Brooks, Daly, Miller, Roper, and
  Wood}]{brooks1996replication}
Brooks A, Daly J, Miller J, et~al (1996) Replication of experimental results in
  software engineering. Technical Report ISERN-96-10, University of
  Strathclyde, Glasgow, UK

\bibitem[{Brooks et~al(2008)Brooks, Roper, Wood, Daly, and Miller}]{Brooks2008}
Brooks A, Roper M, Wood M, et~al (2008) Replication's Role in Software
  Engineering, Springer, pp 365--379. \doi{10.1007/978-1-84800-044-5_14}

\bibitem[{Campbell and Stanley(2015)}]{campbell2015experimental}
Campbell DT, Stanley JC (2015) Experimental and quasi-experimental designs for
  research. Ravenio books

\bibitem[{Carver(2010)}]{carver2010towards}
Carver JC (2010) Towards reporting guidelines for experimental replications: A
  proposal. In: Proceedings of the 1st international workshop on replication in
  empirical software engineering, pp 1--4

\bibitem[{Ciolkowski et~al(2002)Ciolkowski, Shull, and
  Biffl}]{ciolkowski2002family}
Ciolkowski M, Shull F, Biffl S (2002) A family of experiments to investigate
  the influence of context on the effect of inspection techniques. In:
  Proceedings of EASE'02, pp 48--60

\bibitem[{Cruz et~al(2019)Cruz, Bern{\'a}rdez, Dur{\'a}n, Galindo, and
  Ruiz-Cort{\'e}s}]{cruz2019replication}
Cruz M, Bern{\'a}rdez B, Dur{\'a}n A, et~al (2019) Replication of studies in
  empirical software engineering: A systematic mapping study, from 2013 to
  2018. IEEE Access 8:26,773--26,791. \doi{10.1109/ACCESS.2019.2952191}

\bibitem[{Cruz et~al(2021)Cruz, Bern{\'a}rdez, Dur{\'a}n, Guevara, and
  Ruiz-Cort{\'e}s}]{cruzzenodo2021}
Cruz M, Bern{\'a}rdez B, Dur{\'a}n A, et~al (2021) Supplemental material:
  Instantiation of the proposed templates in the multiple case study using
  {CÆSAR LaTeX} template. \doi{10.5281/zenodo.6631976}

\bibitem[{Da~Silva et~al(2014)Da~Silva, Suassuna, Fran{\c{c}}a, Grubb, Gouveia,
  Monteiro, and dos Santos}]{da2014replication}
Da~Silva FQ, Suassuna M, Fran{\c{c}}a ACC, et~al (2014) Replication of
  empirical studies in software engineering research: a systematic mapping
  study. Empirical Software Engineering 19(3):501--557.
  \doi{10.1007/s10664-012-9227-7}

\bibitem[{Dur{\'a}n et~al(1999)Dur{\'a}n, Bern{\'a}rdez, Ruiz-Cort{\'e}s, and
  Toro}]{duran1999requirements}
Dur{\'a}n A, Bern{\'a}rdez B, Ruiz-Cort{\'e}s A, et~al (1999) A requirements
  elicitation approach based in templates and patterns. In: Proceedings of
  WER'99

\bibitem[{Durán et~al(2002)Durán, Corchuelo, Ruiz-Cortés, and
  Toro}]{duran2002supporting}
Durán A, Corchuelo R, Ruiz-Cortés A, et~al (2002) Supporting requirements
  verification using {XSLT}. In: Proceedings of RE'02, pp 165--172,
  \doi{10.1109/ICRE.2002.1048519}

\bibitem[{Fern{\'a}ndez et~al(2020)Fern{\'a}ndez, Graziotin, Wagner, and
  Seibold}]{fernandez2019open}
Fern{\'a}ndez DM, Graziotin D, Wagner S, et~al (2020) Open science in software
  engineering, Springer International Publishing, pp 477--501.
  \doi{10.1007/978-3-030-32489-6_17}

\bibitem[{Fern{\'a}ndez-S{\'a}ez et~al(2016)Fern{\'a}ndez-S{\'a}ez, Genero,
  Caivano, and Chaudron}]{fernandez2016does}
Fern{\'a}ndez-S{\'a}ez AM, Genero M, Caivano D, et~al (2016) Does the level of
  detail of {UML} diagrams affect the maintainability of source code?: a family
  of experiments. Empirical Software Engineering 21(1):212--259.
  \doi{10.1007/s10664-014-9354-4}

\bibitem[{Gamma et~al(1995)Gamma, Helm, Johnson, and
  Vlissides}]{DesignPatterns}
Gamma E, Helm R, Johnson R, et~al (1995) Design Patterns: Elements of Reusable
  Object-Oriented Software. Addison-Wesley

\bibitem[{Garc{\'\i}a et~al(2016)Garc{\'\i}a, Rodr{\'\i}guez-Juan,
  Rodr{\'\i}guez-Guti{\'e}rrez, Rios, and
  Fern{\'a}ndez-Bola{\~n}os}]{garcia2016extraction}
Garc{\'\i}a A, Rodr{\'\i}guez-Juan E, Rodr{\'\i}guez-Guti{\'e}rrez G, et~al
  (2016) Extraction of phenolic compounds from virgin olive oil by deep
  eutectic solvents ({DESs}). Food chemistry 197:554--561.
  \doi{10.1016/j.foodchem.2015.10.131}

\bibitem[{G{\'o}mez et~al(2010)G{\'o}mez, Juristo, and
  Vegas}]{gomez2010replications}
G{\'o}mez OS, Juristo N, Vegas S (2010) Replications types in experimental
  disciplines. In: Proceedings of ESEM'10, p 71–75,
  \doi{10.1145/1852786.1852790}

\bibitem[{G{\'o}mez et~al(2014)G{\'o}mez, Juristo, and
  Vegas}]{gomez2014understanding}
G{\'o}mez OS, Juristo N, Vegas S (2014) Understanding replication of
  experiments in software engineering: A classification. Information and
  Software Technology 56(8):1033--1048. \doi{10.1016/j.infsof.2014.04.004}

\bibitem[{Guevara-Vega et~al(2021)Guevara-Vega, Bern{\'{a}}rdez, Dur{\'{a}}n,
  Qui{\~{n}}a-Mera, Cruz, and Ruiz-Cort{\'{e}}s}]{Guevara-Vega2021}
Guevara-Vega C, Bern{\'{a}}rdez B, Dur{\'{a}}n A, et~al (2021) Empirical
  strategies in software engineering research: A literature survey. In: II
  International Conference on Information Systems and Software Technologies
  (ICI2ST 2021). IEEE Press, Quito, Ecuador

\bibitem[{Carvajal de~la Haza(2016)}]{Sheila2016}
Carvajal de~la Haza S (2016) Copper extraction by brown mustard (brassica
  juncea) plants during vegetative growth in an artificially contaminated soil
  and effect of rhamnolipid application. \url{
  http://hdl.handle.net/11441/50282}, (Final Degree Project). Universidad de
  Sevilla

\bibitem[{Herbold et~al(2017)Herbold, Trautsch, and
  Grabowski}]{herbold2017global}
Herbold S, Trautsch A, Grabowski J (2017) Global vs. local models for
  cross-project defect prediction. Empirical Software Engineering
  22(4):1866--1902. \doi{10.1007/s10664-016-9468-y}

\bibitem[{Itkonen and M{\"a}ntyl{\"a}(2014)}]{itkonen2014test}
Itkonen J, M{\"a}ntyl{\"a} MV (2014) Are test cases needed? replicated
  comparison between exploratory and test-case-based software testing.
  Empirical Software Engineering 19(2):303--342.
  \doi{10.1007/s10664-013-9266-8}

\bibitem[{Jedlitschka et~al(2008)Jedlitschka, Ciolkowski, and
  Pfahl}]{jedlitschka2008reporting}
Jedlitschka A, Ciolkowski M, Pfahl D (2008) Reporting experiments in software
  engineering. In: Guide to advanced empirical software engineering. Springer,
  p 201--228, \doi{10.1007/978-1-84800-044-5_8}

\bibitem[{Juristo and G{\'o}mez(2012)}]{Juristo2012}
Juristo N, G{\'o}mez OS (2012) Replication of software engineering experiments.
  In: Empirical software engineering and verification. Springer, p 60--88,
  \doi{10.1007/978-3-642-25231-0_2}

\bibitem[{Juristo and Moreno(2013)}]{juristo2013basics}
Juristo N, Moreno AM (2013) Basics of software engineering experimentation.
  Springer Science \& Business Media, \doi{10.1007/978-1-4757-3304-4}

\bibitem[{Juristo and Vegas(2003)}]{juristo2003functional}
Juristo N, Vegas S (2003) Functional testing, structural testing and code
  reading: What fault type do they each detect? In: Empirical Methods and
  Studies in Software Engineering. Springer, p 208--232,
  \doi{10.1007/978-3-540-45143-3_12}

\bibitem[{Juristo and Vegas(2009)}]{juristo2009using}
Juristo N, Vegas S (2009) Using differences among replications of software
  engineering experiments to gain knowledge. In: Proceedings of ESEM'09, IEEE,
  pp 356--366, \doi{10.1109/ESEM.2009.5314236}

\bibitem[{Juristo and Vegas(2011)}]{juristo2011role}
Juristo N, Vegas S (2011) The role of non-exact replications in software
  engineering experiments. Empirical Software Engineering 16(3):295--324.
  \doi{10.1007/s10664-010-9141-9}

\bibitem[{Juristo et~al(2012)Juristo, Vegas, Solari, Abrahao, and
  Ramos}]{juristo2012comparing}
Juristo N, Vegas S, Solari M, et~al (2012) Comparing the effectiveness of
  equivalence partitioning, branch testing and code reading by stepwise
  abstraction applied by subjects. In: Proceedings of International Conference
  on Software Testing, Verification and Validation, pp 330--339,
  \doi{10.1109/ICST.2012.113}

\bibitem[{Juristo et~al(2013)Juristo, Vegas, Solari, Abrah{\~a}o, and
  Ramos}]{juristo2013process}
Juristo N, Vegas S, Solari M, et~al (2013) A process for managing interaction
  between experimenters to get useful similar replications. Information and
  Software Technology 55(2):215--225. \doi{10.1016/j.infsof.2012.07.016}

\bibitem[{Kitchenham(2008)}]{kitchenham2008role}
Kitchenham B (2008) The role of replications in empirical software engineering
  a word of warning. Empirical Software Engineering 13(2):219--221.
  \doi{10.1007/s10664-008-9061-0}

\bibitem[{de~Magalh{\~a}es et~al(2015)de~Magalh{\~a}es, da~Silva, Santos, and
  Suassuna}]{de2015investigations}
de~Magalh{\~a}es CV, da~Silva FQ, Santos RE, et~al (2015) Investigations about
  replication of empirical studies in software engineering: A systematic
  mapping study. Information and Software Technology 64:76--101.
  \doi{10.1016/j.infsof.2015.02.001}

\bibitem[{Mendez et~al(2020)Mendez, Graziotin, Wagner, and
  Seibold}]{mendez2020open}
Mendez D, Graziotin D, Wagner S, et~al (2020) Open science in software
  engineering. In: Contemporary Empirical Methods in Software Engineering.
  Springer, p 477--501, \doi{10.1007/978-3-030-32489-6_17}

\bibitem[{Mondal et~al(2018)Mondal, Rahman, Roy, and
  Schneider}]{mondal2018cloned}
Mondal M, Rahman MS, Roy CK, et~al (2018) Is cloned code really stable?
  Empirical Software Engineering 23(2):693--770.
  \doi{10.1007/s10664-017-9528-y}

\bibitem[{Márquez(2018)}]{marquez2018}
Márquez M (2018) Copper extraction by brown mustard (brassica juncea) plants
  during vegetative growth in an artificially contaminated soil and effect of
  rhamnolipid application. \url{ https://hdl.handle.net/11441/132481}, (Final
  Degree Project). Universidad de Sevilla

\bibitem[{Pacheco et~al(2008)Pacheco, L{\'o}pez, Berm{\'u}dez, Abia, Villar,
  and Muriana}]{pacheco2008meal}
Pacheco YM, L{\'o}pez S, Berm{\'u}dez B, et~al (2008) A meal rich in oleic acid
  beneficially modulates postprandial sicam-1 and svcam-1 in normotensive and
  hypertensive hypertriglyceridemic subjects. Journal of Nutritional
  Biochemistry 19(3):200--205. \doi{10.1016/j.jnutbio.2007.03.002}

\bibitem[{Parejo et~al(2016)Parejo, S{\'a}nchez, Segura, Ruiz-Cort{\'e}s,
  Lopez-Herrejon, and Egyed}]{parejo2016multi}
Parejo JA, S{\'a}nchez AB, Segura S, et~al (2016) Multi-objective test case
  prioritization in highly configurable systems: A case study. Journal of
  Systems and Software 122:287--310. \doi{10.1016/j.jss.2016.09.045}

\bibitem[{Plasquy et~al(2021)Plasquy, Florido, Sola-Guirado, and
  Garcia}]{plasquy2021florido}
Plasquy E, Florido MC, Sola-Guirado RR, et~al (2021) Effects of a harvesting
  and conservation method for small producers on the quality of the produced
  olive oil. Agriculture 11. \doi{10.3390/agriculture11050417}

\bibitem[{Reimanis et~al(2014)Reimanis, Izurieta, Luhr, Xiao, Cai, and
  Rudy}]{reimanis2014replication}
Reimanis D, Izurieta C, Luhr R, et~al (2014) A replication case study to
  measure the architectural quality of a commercial system. In: Proceedings of
  ESEM'14, ACM, pp 1--8, \doi{10.1145/2652524.2652581}

\bibitem[{Riaz et~al(2017)Riaz, King, Slankas, Williams, Massacci,
  Quesada-L{\'o}pez, and Jenkins}]{riaz2017identifying}
Riaz M, King J, Slankas J, et~al (2017) Identifying the implied: Findings from
  three differentiated replications on the use of security requirements
  templates. Empirical Software Engineering 22(4):2127--2178.
  \doi{10.1007/s10664-016-9481-1}

\bibitem[{del R{\'\i}o-Ortega et~al(2012)del R{\'\i}o-Ortega, Resinas,
  Dur{\'a}n, and Antonio}]{del2012defining}
del R{\'\i}o-Ortega A, Resinas M, Dur{\'a}n A, et~al (2012) Defining process
  performance indicators by using templates and patterns. In: Proceedings of
  BPM'12, Springer, pp 223--228, \doi{10.1007/978-3-642-32885-5_18}

\bibitem[{del R{\'\i}o-Ortega et~al(2016)del R{\'\i}o-Ortega, Resinas,
  Dur{\'a}n, and Ruiz-Cort{\'e}s}]{del2016using}
del R{\'\i}o-Ortega A, Resinas M, Dur{\'a}n A, et~al (2016) Using templates and
  linguistic patterns to define process performance indicators. Enterprise
  Information Systems 10(2):159--192. \doi{10.1080/17517575.2013.867543}

\bibitem[{del Río(2019)}]{Guijarro2019TFG}
del Río P (2019) Effect of rhamnolipid {JBR}--425 on the development of
  brassica juncea in urban garden soils in {\Sevilla}. \url{
  https://hdl.handle.net/11441/132478}, (Final Degree Project) Universidad de
  Sevilla

\bibitem[{Runeson and H{\"o}st(2009)}]{runeson2009guidelines}
Runeson P, H{\"o}st M (2009) Guidelines for conducting and reporting case study
  research in software engineering. Empirical Software Engineering 14(2):131.
  \doi{10.1007/s10664-008-9102-8}

\bibitem[{Runeson et~al(2012)Runeson, H{\"o}st, Rainer, and
  Regnell}]{runeson2012case}
Runeson P, H{\"o}st M, Rainer A, et~al (2012) Case study research in software
  engineering. In: Guidelines and examples. Wiley Online Library, p 109--126,
  \doi{10.1002/9781118181034.ch8}

\bibitem[{S{\'a}nchez et~al(2014)S{\'a}nchez, Segura, and
  Ruiz-Cort{\'e}s}]{sanchez2014comparison}
S{\'a}nchez AB, Segura S, Ruiz-Cort{\'e}s A (2014) A comparison of test case
  prioritization criteria for software product lines. In: Proceedings of
  ICST'14, pp 41--50, \doi{10.1109/ICST.2014.15}

\bibitem[{Santos et~al(2018)Santos, G{\'o}mez, and
  Juristo}]{santos2018analyzing}
Santos A, G{\'o}mez OS, Juristo N (2018) Analyzing families of experiments in
  {\SE}: a systematic mapping study. IEEE Transactions on Software Engineering
  46(5):566--583. \doi{10.1109/TSE.2018.2864633}

\bibitem[{Santos et~al(2019)Santos, do~Carmo~Machado, de~Almeida, Siegmund, and
  Apel}]{santos2019comparing}
Santos AR, do~Carmo~Machado I, de~Almeida ES, et~al (2019) Comparing the
  influence of using feature-oriented programming and conditional compilation
  on comprehending feature-oriented software. Empirical Software Engineering
  24(3):1226--1258. \doi{10.1007/s10664-018-9658-x}

\bibitem[{dos Santos et~al(2022)dos Santos, de~Almeida, and
  Ahmed}]{dos2022investigating}
dos Santos DA, de~Almeida ES, Ahmed I (2022) Investigating replication
  challenges through multiple replications of an experiment. Information and
  Software Technology p 106870. \doi{10.1016/j.infsof.2022.106870}

\bibitem[{Scanniello et~al(2015)Scanniello, Marcus, and
  Pascale}]{scanniello2015link}
Scanniello G, Marcus A, Pascale D (2015) Link analysis algorithms for static
  concept location: an empirical assessment. Empirical Software Engineering
  20(6):1666--1720. \doi{10.1007/s10664-014-9327-7}

\bibitem[{Segura et~al(2017)Segura, Dur{\'a}n, Troya, and
  Cort{\'e}s}]{segura2017template}
Segura S, Dur{\'a}n A, Troya J, et~al (2017) A template-based approach to
  describing metamorphic relations. In: Proceedings of International Workshop
  on Metamorphic Testing, pp 3--9, \doi{10.1109/MET.2017.3}

\bibitem[{Shepperd et~al(2018)Shepperd, Ajienka, and
  Counsell}]{shepperd2018role}
Shepperd M, Ajienka N, Counsell S (2018) The role and value of replication in
  empirical software engineering results. Information and Software Technology
  99:120--132. \doi{10.1016/j.infsof.2018.01.006}

\bibitem[{Shull et~al(2002)Shull, Basili, Carver, Maldonado, Travassos,
  Mendon{\c{c}}a, and Fabbri}]{shull2002replicating}
Shull F, Basili V, Carver J, et~al (2002) Replicating software engineering
  experiments: addressing the tacit knowledge problem. In: Proceedings of
  ISESE'02, pp 7--16, \doi{10.1109/ISESE.2002.1166920}

\bibitem[{Shull et~al(2008)Shull, Carver, Vegas, and Juristo}]{shull2008role}
Shull F, Carver JC, Vegas S, et~al (2008) The role of replications in empirical
  software engineering. Empirical Software Engineering 13(2):211--218.
  \doi{10.1007/s10664-008-9060-1}

\bibitem[{Solari(2013)}]{solari2013identifying}
Solari M (2013) Identifying experimental incidents in software engineering
  replications. In: Proceedings of ESEM'13, IEEE, pp 213--222,
  \doi{10.1109/ESEM.2013.26}

\bibitem[{Solari and Vegas(2006)}]{solari2006classifying}
Solari M, Vegas S (2006) Classifying and analysing replication packages for
  software engineering experimentation. In: Proceedings of WSESE'06 Amsterdam,
  Paises Bajos, pp 19--24

\bibitem[{Solari et~al(2017)Solari, Vegas, and Juristo}]{solari2017content}
Solari M, Vegas S, Juristo N (2017) Content and structure of laboratory
  packages for software engineering experiments. Information and Software
  Technology 97:64--79. \doi{10.1016/j.infsof.2017.12.016}

\bibitem[{Spence and Stanley(2016)}]{spence2016prediction}
Spence JR, Stanley DJ (2016) Prediction interval: What to expect when you’re
  expecting… a replication. PloS one 11(9):e0162,874.
  \doi{10.1371/journal.pone.0162874}

\bibitem[{Vegas et~al(2020)Vegas, Riofr{\'\i}o, Marcos, and
  Juristo}]{vegas2020mis}
Vegas S, Riofr{\'\i}o P, Marcos E, et~al (2020) On (mis) perceptions of testing
  effectiveness: an empirical study. Empirical Software Engineering
  25:2844--2896. \doi{10.1007/s10664-020-09805-y}

\bibitem[{Von~Alan et~al(2004)Von~Alan, March, Park, and Ram}]{von2004design}
Von~Alan RH, March ST, Park J, et~al (2004) Design science in information
  systems research. MIS quarterly 28(1):75--105. \doi{10.2307/25148625}

\bibitem[{Wieringa(2014)}]{38631e0608b54d4299d5707f3a78debf}
Wieringa RJ (2014) Design science methodology for information systems and
  software engineering. Springer

\bibitem[{Wohlin et~al(2012)Wohlin, Runeson, Höst, Ohlsson, Regnell, and
  Wesslén}]{wohlin:experimentation}
Wohlin C, Runeson P, Höst M, et~al (2012) Experimentation in software
  engineering: an introduction. Springer

\bibitem[{Yin et~al(2003)}]{yin2003design}
Yin RK, et~al (2003) Design and methods. Case study research. Sage
  Publications: Thousand Oaks

\end{thebibliography}

\appendix



\section{Online Resources}


The specification of the replications and their corresponding changes selected in the multiple case study 
are available at the Zenodo repository \cite{cruzzenodo2021}.

The \caesar tool is available at \caesarurl.

\end{document}